\documentclass[conference]{IEEEtran}
\IEEEoverridecommandlockouts
\usepackage{cite}
\usepackage{amsmath,amssymb,amsfonts}
\usepackage{algorithmic}
\usepackage{graphicx}
\usepackage{textcomp}
\usepackage{tikz}
\usetikzlibrary{shapes,arrows}
\usepackage{float}
\usepackage{xcolor}
\def\BibTeX{{\rm B\kern-.05em{\sc i\kern-.025em b}\kern-.08em
    T\kern-.1667em\lower.7ex\hbox{E}\kern-.125emX}}

\usepackage{flushend}

\newcommand{\figref}[1]{Fig. \ref{#1}}

\def\anonymize{0}

\if\anonymize1
	\newcommand{\blindtext}[1]{\phantom{#1}}
	\else 
	\newcommand{\blindtext}[1]{#1}
\fi

\onecolumn

\begin{document}

\title{SNN Architecture for Differential Time Encoding Using Decoupled Processing Time}

\author{\IEEEauthorblockN{\blindtext{Daniel Windhager}}
\IEEEauthorblockA{\blindtext{\textit{Intelligent Wireless Systems}} \\
\blindtext{\textit{Silicon Austria Labs}}\\
\blindtext{Linz, Austria} \\
\blindtext{daniel.windhager@silicon-austria.com}}
\and
\IEEEauthorblockN{\blindtext{Bernhard A. Moser$^*$}}\thanks{\blindtext{$^*$double affiliation: Software Competence Center Hagenberg (SCCH), 4232 Hagenberg, Austria}}
\IEEEauthorblockA{\blindtext{\textit{Institute of Signal Processing}} \\
\blindtext{\textit{Johannes Kepler University}}\\
\blindtext{Linz, Austria} \\
\blindtext{bernhard.moser@jku.at}}
\and
\IEEEauthorblockN{\blindtext{Michael Lunglmayr}}
\IEEEauthorblockA{\blindtext{\textit{Institute of Signal Processing}} \\
\blindtext{\textit{Johannes Kepler University}}\\
\blindtext{Linz, Austria} \\
\blindtext{michael.lunglmayr@jku.at}}
}

\maketitle

\begin{abstract}
Spiking neural networks (SNNs) have gained  attention in recent years due to their ability to handle sparse and event-based data better than regular artificial neural networks (ANNs). Since  the structure of SNNs is less suited for typically used accelerators such as GPUs than conventional ANNs, there is a demand for custom hardware accelerators for processing SNNs. In the past, the main focus was on platforms that resemble the structure of multiprocessor systems. In this work, we propose a lightweight neuron layer architecture that allows network structures to be directly mapped onto digital hardware.
Our approach is based on differential time coding of spike sequences and the decoupling of processing time and spike timing that allows the SNN to be processed on different hardware platforms.
We present synthesis and performance results showing that this architecture can be implemented for networks of more than $1000$ neurons with high clock speeds on a State-of-the-Art FPGA. We furthermore show results on the robustness of our approach to quantization. These results demonstrate that high-accuracy inference can be performed with bit widths as low as $4$.
\end{abstract}


\section{Introduction}
In recent years artificial, intelligence has revolutionized countless fields, ranging from machine vision and recognition to generative AI. While the currently available architectures, such as Convolutional Neural Networks (CNNs) and Transformers, are undoubtedly powerful, the size of these networks is constantly increasing, and they require more and more energy and computing power~\cite{HuChu2021}.

This naturally led to other architectures being researched, including spiking neural networks (SNNs)~\cite{MAASS19971659}. 
While common artificial neural networks process real vectors, the neurons in an SNN process sequences of spikes which are also referred to as spike trains. In a spike train $S$, the information is encoded in the position of the spikes, i.e. their spike times $t_k$, as well as in the amplitudes $a_{k}$ of the spikes, i.e.,
\begin{equation}
  S = \sum_{k}a_{k}\delta(t-t_{k}).
\label{eq:spike_train}
\end{equation}
For ease of presentation, we assume that the spike times $t_k$ are sorted in ascending order.  
Since the time of occurrence of the spikes is very important, many architectures only re-compute the output when an actual spike occurs, which is why these architectures are also sometimes called ``event-driven'' or ``event-based''. This event-driven approach of only computing during a spike event already significantly influences the power consumption and the required computation effort, especially for sparse spike trains. Unfortunately, this event-driven approach can often not be efficiently mapped to multi-processor and GPU architectures, which is why many custom solutions, such as IBM's TrueNorth \cite{truenorth} or Intel's Loihi \cite{loihi}, were specifically designed to accelerate these SNN architectures. 

This work describes a digital architecture for efficiently processing time-discrete spike trains in digital hardware. When using a spiking neural network, the encoding of data into spike trains is of great importance. 
In contrast to the state of the art, which mainly uses  
so-called rate coding that represents a data value by the number of ones in a 
spike train per time interval, in this work we use delta encoding.
 When sampling analog signals, this encoding can be efficiently obtained by using 
threshold-based sampling where a spike is triggered when the amplitude change of a 
signal surpasses a certain threshold~\cite{Moser2017SimilarityRF}. 
This also motivated us to use differential time encoding of the spike times $t_k$ by storing only the time information $\Delta t_k = t_k - t_{k-1}$, see~\cite{aisys}. Our SNN architecture aims at differential encoding of both time and amplitude.

Our main contributions are: a) the proposal of a novel low-complexity architecture for efficient processing of full layers of SNNs using a differential time encoding of spike trains, decoupling spike time and processing time, b) inference performance results showing the superiority on the MNIST~\cite{mnist} dataset to a recent state-of-the-art solution,
and, c) performance results demonstrating its robustness to quantization errors.

\section{Architecture description}
\subsection{Network Structure}
Many architectures that are designed to accelerate SNNs use a structure that is similar to modern multiprocessor systems, i.e. they typically include a number of neuron processors which are placed in a row-column layout with multiple connections in-between (e.g. \cite{TCASSPIKEMNIST21, energy_efficient_neuromorph_arch, noc}).
This commonly results in fewer but more powerful neurons being used to ensure that the amount of routing between neurons is not the main factor in resource consumption. 

\begin{figure}[h]
\centering
\includegraphics[width=0.9\linewidth]{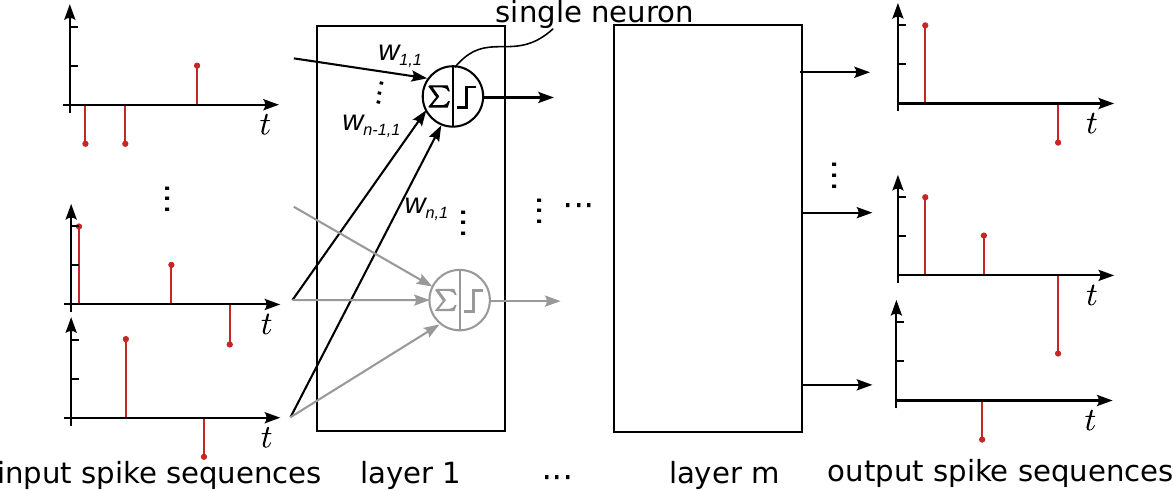}
\caption{SNN structure via its layers.}
\label{fig:network_structure}
\end{figure}

In this work, we describe
a different approach, i.e. interconnecting a large number of low-complexity, but still powerful neurons.
This allows directly connecting the neurons and thus implementing a large number of neurons on a hardware platform without the communication overheads commonly found in multiprocessor-like structures used for neuron processing.  
For this, the neurons are arranged in layers, where each neuron is only connected to the neurons in the previous and following layer, but not to neurons in the same layer, which is schematically shown in \figref{fig:network_structure}. 
In this work, we use a fully connected network. One can, however, easily cut synapse connections by setting the corresponding synapse weights to zero.
Due to the low complexity of the neurons in our architecture, each neuron of a desired SNN structure is mapped to a physical neuron in digital hardware. The communication between the neurons then naturally happens via the connection signals between the neurons. As we will demonstrate below, this allows implementing over $1000$ neurons even in moderately-sized FPGAs. In case a network does not fully fit into a given hardware platform, e.g. an FPGA, the structure easily allows implementing a subnetwork that is then used multiple times while storing intermediate spike data in memory. 

\subsection{Spike encoding}

In addition to proposing an alternative SNN processing architecture, the design proposed in this work also deviates from the commonly used way to represent spikes. It is quite obvious that simply storing the spikes for every single timestep as given in~\eqref{eq:spike_train} lacks efficiency. Therefore, many designs (see \cite{noc,darwin,snava}) use the Address-Event Representation (AER) format to represent spikes, which consists of an identifier for a neuron, like a row-column tuple, as well as a timestamp that is used to indicate when the spike occurred. A major disadvantage of the AER is that the timestamp is usually given in absolute time, meaning the number of bits needed to represent the timestamp grows logarithmically with the maximum number of cycles that the design can be active before being reset. Contrary to the AER, we do not store absolute time, but rather the delta time (given in cycles of arbitrary precision) between the spikes. Fortunately, converting from spikes given in absolute time to spikes using delta time is trivial. For a vector of non-decreasing spike-times $t_{k}$, the corresponding vector of spikes in delta time can be computed as $\Delta t_k = t_k -t_{k-1}$.

\begin{figure}
	\centering
	\includegraphics[width=0.9\linewidth]{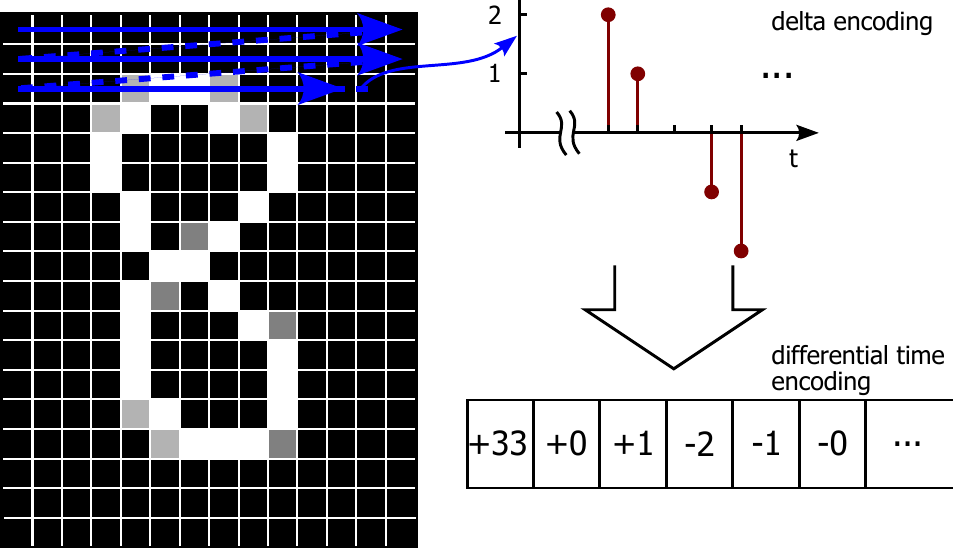}
	\caption{Delta encoding of image represented via differential time encoding}
	\label{fig:imagetospike}
\end{figure}
Fig.~\ref{fig:imagetospike} schematically shows the encoding of an example image via delta encoding that is then converted into a differential time representation. First, the image is converted to a spike train where the spike amplitude reflects the brightness-change from one pixel to the next (the  pixel index is interpreted as time here).
Second, the spike sequence is converted to a differential time format as shown on the bottom right in Fig.~\ref{fig:imagetospike}. It uses a sign-magnitude format to encode the sign of the spike and the time difference to the last spike. Using a sign-magnitude format conveniently allows encoding spikes of amplitudes larger than one, as this format allows having $\pm 0$. A spike of amplitude $a\in\mathbb{N}_{0}$ is represented by a spike of amplitude $1$ followed by $a-1$ spikes with time difference of $0$, e.g. a spike of height $2$ can be encoded by a positive differential time followed by a $+0$ as shown in Fig.~\ref{fig:imagetospike}.

Apart from being more efficient for spikes that occur with little time in between, it also allows for (almost) arbitrary precision regarding the number of bits for representing the time delta. Conventionally, we would need to choose the number of bits to fit the maximum delay
between two spikes, however this can be circumvented if a new virtual spike is introduced (we reserved the all-ones bit pattern for this), which we use as a timer overflow spike. Similar to a regular spike it is propagated through the network, but the only information it carries is the passing of time.
No neuron potential will be adjusted and no new spikes will be generated, except for more overflow spikes in later stages of the network. This leads to a tradeoff between the bit size available for spikes and the number of spikes in the system: with less bits, more overflow spikes will be generated, while more bits will decrease the number of necessary overflow spikes, at the cost of higher resource usage. Furthermore, since the connections between the neurons are fixed during the development stage, no further message passing overhead is necessary as the spikes will simply travel from each neuron in layer $l$ to every neuron in layer $l+1$.
Using such a data encoding allows decoupling the processing time from the spike time as the outputs of the network will be vectors of differentially time-encoded spike data. The clock cycles to process these data vectors are completely decoupled from the timing encoded in these vectors (see below for more details).
 
\subsection{Basic Neuron Function}
The neuron design presented in this work is based on the commonly used Leaky-Integrate and Fire (LIF) neuron with a few 
adaptations~\cite{moser2023quantization}. An integral part of the LIF neuron is the neuron potential inherent to each neuron, which is continually updated during inference. The rules for updating the neuron potential can be summarized as
\begin{equation}
  \label{eq:lif_potential_update}
  P_{k} = r_{f}\left(P_{k-1}\beta^{\Delta{}t} + \sum_{i}w_{i}s_{i}\right)
\end{equation}
where the function $P_{k}$ is the neuron potential during update cycle $k$, $\beta \in [0,1]$ is the decay factor, $\Delta{}t$ is the time since the last update, and $w_{i}$ and $s_{i}$ are the weights and spike amplitudes for each input synapse, respectively. 
 The function  $r_{f}$ is responsible for resetting the potential if the neuron produces an output spike. This can be described as
\begin{equation}
  \label{eq:neuron_potential_reset}
  r_{f}(P) = \begin{cases}
               P,        &\text{if } \theta_{\mbox{low}} < P < \theta_{\mbox{high}}\\
               P-\theta_{\mbox{low}},  &\text{if } P\leq{}\theta_{\mbox{low}}\\
               P-\theta_{\mbox{high}}, &\text{if } P\geq{}\theta_{\mbox{high}},
  \end{cases}
\end{equation}
where, $\theta_{\mbox{high}}$ and $\theta_{\mbox{low}}$ are the thresholds to trigger positive and negative spikes at the neuron's output, respectively ($\theta_{\mbox{low}}$ is optional).
In this work, we assume that $\beta$ is a power of two, i.e. $\beta = 2^{-i}, {i \in \mathbb{N}}$ (for the results presented below we used $\beta = 0.5$). 
This simplifies the neuron core, as described below, as 
$2^{-(i \Delta t)}$ can be efficiently calculated via shift operations. 
Although this might seem like a restriction of the neuron model at first glance, the following arguments show that, practically, it does not restrict a neuron's power.
Assuming one wants to use an arbitrary decay $\beta$ that is not a power of two, then one could set 
$\beta^{\Delta t} = {2^{-i}}^{\lambda \Delta t}$
to calculate the time conversion factor $\lambda = \text{log}(\beta)/\text{log}(2^{-i})$.
Therefore, globally scaling all spike times (and consequently also the delta times between the spikes) with this factor $\lambda$ causes the neurons to act as if an arbitrary decay was used, while $\beta$ can still be set to a power of two. However, as our experimental results shown below demonstrate, one can achieve excellent results without adjusting the time values and keeping $\beta$ as a power of two. For the investigated problem, using $\beta=0.5$, the neuron weights provided a large enough solution space for the SNN training algorithm to find good results.

\subsection{Hardware Design}
To process spike trains with the differential temporal coding described above, we need to ensure that causality is preserved, i.e. spikes are processed in the correct order, especially when different spike trains are combined at one neuron. Since a neuron typically has multiple input synapses, each of which can transport spikes to the neuron, one must ensure the correct temporal relationship between the different incoming spike trains (the encoded timing differences describe the timing within a spike train, but of course not between spike trains).

\begin{figure}[h]
  \centering
  \includegraphics[width=0.9\linewidth]{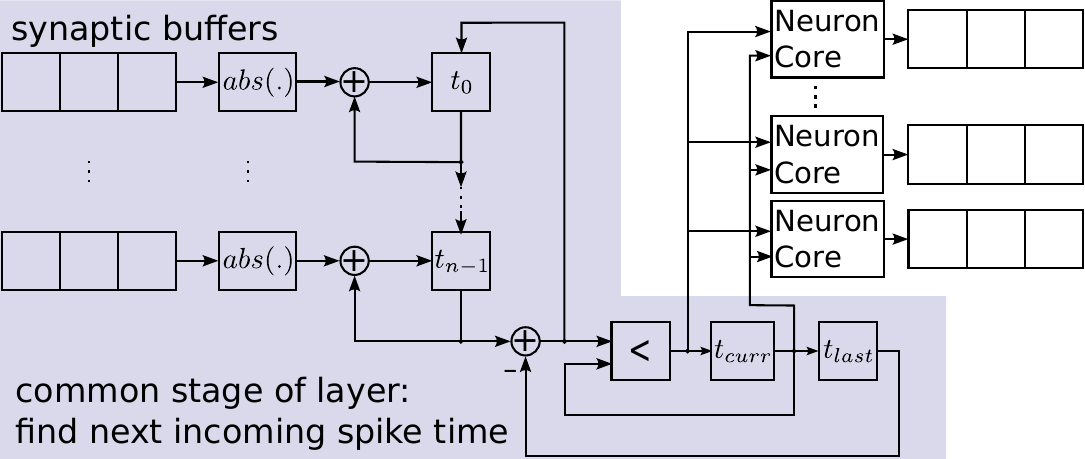}
  \caption{Architecture of SNN neuron layer}
  \label{fig:hw_block_diagram}
\end{figure}

\begin{figure}[h]
	\centering
	\includegraphics[width=0.9\linewidth]{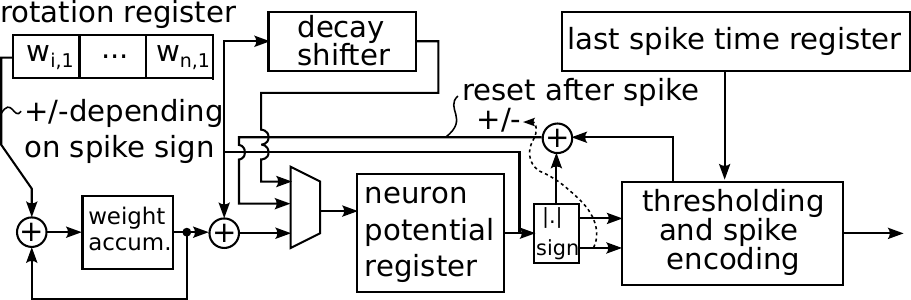}
	\caption{Architecture of SNN neuron layer}
	\label{fig:NeuronCore}
\end{figure}

Fig.~\ref{fig:hw_block_diagram} graphically shows the structure of a neuron layer.
By using a signed-magnitude representation, one can easily recover the sign and time difference from the incoming synapse buffers. In order to properly 
relate spike trains to each other, for each input synapse we use an integrator for converting the difference time to absolute time.
Because of the used overflow spikes, this time integrator is naturally limited in its bit width: each time a timer overflow at the output of a neuron is triggered, the integrators are reduced by the corresponding time of the overflow spike (as is also the spike time register of a layer), thus preventing the bit width of the integrators growing to infinity. As we are using fully connected layers
we process all integrated times of the incoming spikes of a layer together
to (based on the current system spike time register $t_{curr}$) find the time of the next incoming spike. As all incoming spikes are connected to all neurons of a layer, this can be done once per layer, since only the weights of the neurons differ.
For this, the times $t_0, \ldots, t_{n-1}$ of a layer are rotated $n$ times to find their minimum, i.e. the time the next input spike occurs. This rotation is done via the system clock of the circuit, thus, the time of the spike sequences (spike time) is decoupled from this system clock.
Note that this way  one can interpret the spike times encoded in the differential representation as being stopped while a neuron layer does its processing. Obviously, for a real-time implementation, where a sampling unit produces the spike trains, one has to choose an architecture with a large enough processing frequency to fulfill the real-time constraints.  
After the next spike-time is found, it is handed over to the neuron cores of all the neurons of a layer. Fig.~\ref{fig:NeuronCore} schematically shows the structure of a neuron core. Its rotational register containing the synapse weights is rotated in parallel to the rotation of the time register of the common stage of a layer, allowing to add the weights of the synapses with the correct spike sign together. Whenever a new minimum time is found, the weight accumulator is reset. If another spike with the same time as the current minimum is found, the corresponding weight of that spike is simply added to the accumulator.
This part of the architecture shows an important element for the quantization
effects within the architecture.
If the bit width of the weight accumulator and the neuron potential is sufficiently large, no further quantization effects can occur due to the arithmetic operations within our architecture, as the spikes are of integer amplitude.
After this, the neuron potential is decayed via a shifter (as a power of two is used for $\beta$) and the accumulated weights are added to the decayed potential. If the potential exceeds the threshold (the architecture also allows triggering of negative spikes by comparing with a negative threshold), an outgoing spike is generated and encoded using the differential time format based on the last time a spike was triggered (the last spike time register is also reduced when an overflow spike is triggered limiting its bit width as well).

After a spike triggering event, the spike potential is reset accordingly. As reset mechanism we subtract the threshold with the sign of the triggered spike. As the architecture can trigger spikes with zero differential time, this reset mechanism can be seen as an extended variant of reset-by-subtraction leading to a modulo operation referred to as {\it reset-to-mod} due to~\cite{moser2023quantization, moser2023spiking}. 

\section{Simulation and Synthesis Results}
We tested our SNN architecture for the MNIST dataset \cite{mnist}. We used a fully connected network of $784$ inputs connected to a 
hidden layer composed of $1000$ neurons that are in turn connected to $10$ output neurons (i.e. each neuron of the hidden layer has $784$ synapses, each neuron of the output layer has $1000$ synapses). Each of the output neurons corresponds to a digit $\{0,\ldots,9\}$. The neuron outputting the most spikes is selected to classify the digit number (we only use a positive threshold of one for all neurons in the network and consequently a neuron can only trigger positive spikes).
To obtain the following results we used SNNTorch \cite{SNNTorch} and closely followed Tutorial $5$ \cite{snnTorchSurrogate} using the \emph{LeakySurrogate} neuron on the MNIST dataset where the pixel values were normalized to $[0,1]$ as in the tutorial, using $100$ training epochs. However, different to the tutorial we use delta encoding of the spikes as shown in Fig.~\ref{fig:imagetospike} using a threshold value of $0.05$. The corresponding spike sequence of an image is fed into the network in parallel $28$ times.
\\\\
We also investigated the influence of the quantization of the weights on the inference performance of our architecture. As described above, one of the key advantages is that only integer multiples of synapse weights are added to a neuron's potential. Tab.~\ref{tab:quant} shows these inference results when quantizing the weights and running the SNNs with the quantized weights. As the results displayed in Tab.~\ref{tab:quant} show, the architecture described in this work allows obtaining a good inference performance with bit widths as low as $4$ bits.
\begin{table}[t]
	\centering
	\caption{Comparison with SNN implementation for MNIST reported in literature~\label{tab:quant}}
	\setlength\tabcolsep{3pt} 
	\begin{tabular}{|c|c|}
		\hline	
		bit width & inference accuracy \\
		\hline	
		4 &$95.84\%$ \\
		5 &$96.90\%$ \\
		6 &$97.00\%$ \\
		7 &$97.07\%$ \\
		8 &$97.05\%$ \\
		9 &$97.08\%$ \\
		\hline
	\end{tabular}
\end{table}

Tab.~\ref{tab:Comparison} shows a comparison of the synthesis results of the implementation described in this work and a recently published work which described an SNN implementation for the MNIST digit recognition dataset.
\begin{table}[t]
	\centering
	\caption{ Comparison with SNN implementation for MNIST reported in literature\label{tab:Comparison}}
	\setlength\tabcolsep{3pt} 
	\begin{tabular}{|c|c|c|}
    \hline
                            & \centering\cite{TCASSPIKEMNIST21} &  this work      \\\hline
    FPGA/Board              &          Virtex~7 VC707           &  Zynq ZCU102    \\
    Max.~clock frequ. (MHz) &           $\approx 100$           &  $\approx 365$  \\
    Clock cycles (avg.)     &           not reported            &  435368         \\
    Correctly classified    &            $92.92\,\%$            &  $97.00\%$      \\
    Class. images/second    &           $\approx 320$           &  $\approx 838$  \\
    Energy                  &           $5.04$mJ/image          &  $5.46$mJ/image \\
    CLBs                    &           not reported            &  108894         \\
    Registers               &           not reported            &  57659          \\
    \hline
	\end{tabular}
\end{table}
As one can see from the results, the design proposed in this work achieves very good results for the MNIST dataset while still being small enough to be easily implemented on currently available FPGAs. Furthermore, our architecture requires no multipliers, yet still achieves $97\%$ accuracy on the MNIST test set, while only using 6-bit fixed point values for the weights, as compared to the 16-bit floating point values that~\cite{TCASSPIKEMNIST21} uses when using an FPGA of comparable size. While there are other designs available for comparison (such as \cite{noc, snava, minitaur}), \cite{TCASSPIKEMNIST21} already includes a comparison outperforming many of these architectures.

\section{Conclusion}
This work introduced a hardware design that allows for efficient processing of differential time encoded spikes, using no multipliers and requiring only low bit widths while still achieving $97\%$ accuracy on the MNIST test dataset. By decoupling processing time and spike time, the architecture allows efficiently processing spike trains.

\section*{\blindtext{Acknowledgment}}
\blindtext{The research reported in this paper has been partly funded by BMK, BMDW, and the State of Upper Austria in the frame of SCCH, part of the COMET Programme managed by FFG. This work has been supported by Silicon Austria Labs (SAL), owned by the Republic of Austria, the Styrian Business Promotion Agency (SFG), the federal state of Carinthia, the Upper Austrian Research (UAR), and the Austrian Association for the Electric and Electronics Industry (FEEI).}

\bibliographystyle{IEEEtran}
\bibliography{bibliography.bib}

\begin{thebibliography}{10}
\providecommand{\url}[1]{#1}
\csname url@samestyle\endcsname
\providecommand{\newblock}{\relax}
\providecommand{\bibinfo}[2]{#2}
\providecommand{\BIBentrySTDinterwordspacing}{\spaceskip=0pt\relax}
\providecommand{\BIBentryALTinterwordstretchfactor}{4}
\providecommand{\BIBentryALTinterwordspacing}{\spaceskip=\fontdimen2\font plus
\BIBentryALTinterwordstretchfactor\fontdimen3\font minus
  \fontdimen4\font\relax}
\providecommand{\BIBforeignlanguage}[2]{{%
\expandafter\ifx\csname l@#1\endcsname\relax
\typeout{** WARNING: IEEEtran.bst: No hyphenation pattern has been}%
\typeout{** loaded for the language `#1'. Using the pattern for}%
\typeout{** the default language instead.}%
\else
\language=\csname l@#1\endcsname
\fi
#2}}
\providecommand{\BIBdecl}{\relax}
\BIBdecl

\bibitem{HuChu2021}
\BIBentryALTinterwordspacing
X.~Hu, L.~Chu, J.~Pei, W.~Liu, and J.~Bian, ``Model complexity of deep
  learning: A survey,'' \emph{Knowl. Inf. Syst.}, vol.~63, no.~10, p.
  2585–2619, oct 2021. [Online]. Available:
  \url{https://doi.org/10.1007/s10115-021-01605-0}
\BIBentrySTDinterwordspacing

\bibitem{MAASS19971659}
W.~Maass, ``Networks of spiking neurons: The third generation of neural network
  models,'' \emph{Neural Networks}, vol.~10, no.~9, pp. 1659--1671, 1997.

\bibitem{truenorth}
F.~Akopyan, J.~Sawada, A.~Cassidy, R.~Alvarez-Icaza, J.~Arthur, P.~Merolla,
  N.~Imam, Y.~Nakamura, P.~Datta, G.-J. Nam, B.~Taba, M.~Beakes, B.~Brezzo,
  J.~B. Kuang, R.~Manohar, W.~P. Risk, B.~Jackson, and D.~S. Modha,
  ``Truenorth: Design and tool flow of a 65 mw 1 million neuron programmable
  neurosynaptic chip,'' \emph{IEEE Trans. on Computer-Aided Design of
  Integrated Circuits and Systems}, vol.~34, no.~10, pp. 1537--1557, 2015.

\bibitem{loihi}
M.~Davies, A.~Wild, G.~Orchard, Y.~Sandamirskaya, G.~A.~F. Guerra, P.~Joshi,
  P.~Plank, and S.~R. Risbud, ``Advancing neuromorphic computing with loihi: A
  survey of results and outlook,'' \emph{Proceedings of the IEEE}, vol. 109,
  no.~5, pp. 911--934, 2021.

\bibitem{Moser2017SimilarityRF}
B.~A. Moser, ``Similarity recovery from threshold-based sampling under general
  conditions,'' \emph{IEEE Trans. on Signal Processing}, vol.~65, pp.
  4645--4654, 2017.

\bibitem{aisys}
M.~Lunglmayr, G.~Lindorfer, and B.~Moser, ``Robust and efficient bio-inspired
  data-sampling prototype for time-series analysis,'' in \emph{DEXA
  2021}.\hskip 1em plus 0.5em minus 0.4em\relax Springer International
  Publishing, 2021, pp. 119--126.

\bibitem{mnist}
\BIBentryALTinterwordspacing
Y.~LeCun, C.~Cortes, and C.~Burges, ``{MNIST Handwritten Digit Database},''
  2010. [Online]. Available: \url{http://yann.lecun.com/exdb/mnist}
\BIBentrySTDinterwordspacing

\bibitem{TCASSPIKEMNIST21}
S.~Li, Z.~Zhang, R.~Mao, J.~Xiao, L.~Chang, and J.~Zhou, ``{A Fast and
  Energy-Efficient SNN Processor With Adaptive Clock/Event-Driven Computation
  Scheme and Online Learning},'' \emph{IEEE Trans. on Circuits and Systems I:
  Regular Papers}, vol.~68, no.~4, pp. 1543--1552, 2021.

\bibitem{energy_efficient_neuromorph_arch}
\BIBentryALTinterwordspacing
Q.~Wang, Y.~Li, B.~Shao, S.~Dey, and P.~Li, ``Energy efficient parallel
  neuromorphic architectures with approximate arithmetic on fpga,''
  \emph{Neurocomputing}, vol. 221, pp. 146--158, 2017. [Online]. Available:
  \url{https://www.sciencedirect.com/science/article/pii/S0925231216311213}
\BIBentrySTDinterwordspacing

\bibitem{noc}
H.~Fang, A.~Shrestha, D.~Ma, and Q.~Qiu, ``Scalable noc-based neuromorphic
  hardware learning and inference,'' in \emph{IJCNN}, 2018, pp. 1--8.

\bibitem{darwin}
D.~Ma, J.~Shen, Z.~Gu, M.~Zhang, X.~Zhu, X.~Xu, Q.~Xu, Y.~Shen, and G.~Pan,
  ``Darwin: A neuromorphic hardware co-processor based on spiking neural
  networks,'' \emph{Journal of Systems Architecture}, vol.~77, pp. 43--51,
  2017.

\bibitem{snava}
A.~Sripad, G.~Sanchez, M.~Zapata, V.~Pirrone, T.~Dorta, S.~Cambria, A.~Marti,
  K.~Krishnamourthy, and J.~Madrenas, ``Snava—a real-time multi-fpga
  multi-model spiking neural network simulation architecture,'' \emph{Neural
  Networks}, vol.~97, pp. 28--45, 2018.

\bibitem{moser2023quantization}
B.~A. Moser and M.~Lunglmayr, ``Quantization in spiking neural networks,''
  \emph{arXiv}, 2023.

\bibitem{moser2023spiking}
------, ``Spiking neural networks in the alexiewicz topology: A new perspective
  on analysis and error bounds,'' \emph{arXiv}, 2023.

\bibitem{SNNTorch}
J.~K. Eshraghian, M.~Ward, E.~O. Neftci, X.~Wang, G.~Lenz, G.~Dwivedi,
  M.~Bennamoun, D.~S. Jeong, and W.~D. Lu, ``Training spiking neural networks
  using lessons from deep learning,'' \emph{Proceedings of the IEEE}, vol. 111,
  no.~9, pp. 1016--1054, 2023.

\bibitem{snnTorchSurrogate}
\BIBentryALTinterwordspacing
J.~K. Eshraghian, ``{snnTorch Tutorial 5 - Training Spiking Neural Networks
  with snntorch}.'' [Online]. Available:
  \url{https://snntorch.readthedocs.io/en/latest/tutorials/tutorial\_5.html}
\BIBentrySTDinterwordspacing

\bibitem{minitaur}
D.~Neil and S.-C. Liu, ``Minitaur, an event-driven fpga-based spiking network
  accelerator,'' \emph{IEEE Transactions on Very Large Scale Integration (VLSI)
  Systems}, vol.~22, no.~12, pp. 2621--2628, 2014.

\end{thebibliography}

\end{document}